\documentclass[12pt]{amsart}

\usepackage{amsmath,amssymb,lscape}
\usepackage[all]{xy}
\usepackage{enumerate}

\newtheorem{theorem}{Theorem}[section]
\newtheorem{proposition}[theorem]{Proposition}
\newtheorem{lemma}[theorem]{Lemma}

\theoremstyle{definition}

\newtheorem{example}[theorem]{Example}

\newtheorem{problem}[theorem]{Problem}

\theoremstyle{remark}
\newtheorem{remark}[theorem]{Remark}

\numberwithin{equation}{section}

\def\DJ{{\hbox{D\kern-.8em\raise.15ex\hbox{--}\kern.35em}}}
\def\DJo{$\;$\kern-.4em
    \hbox{D\kern-.8em\raise.15ex\hbox{--}\kern.35em koori\'c}}

\def\bQ{{\mathbb {Q}}}

\def\bR{{\mathbb {R}}}
\def\bC{{\mathbb {C}}}

\def\bH{{\mathbb {H}}}

\def\pA{{\mathcal A}}
\def\pB{{\mathcal B}}
\def\pF{{\mathcal F}}
\def\pE{{\mathcal E}}

\def\pQ{{\mathcal Q}}

\def\pU{{\mathcal U}}

\def\pO{{\mathcal O}}

\def\Tr{\mbox{\rm Tr\,}}

\def\Ort{{\mbox{\rm O}}}
\def\Sp{{\mbox{\rm Sp}}}
\def\U{{\mbox{\rm U}}}

\renewcommand{\subjclassname}{\textup{2000} Mathematics Subject
Classification} 

\begin{document}

\title[ortho-, uni- and qu-stochastic matrices]
{On orthostochastic, unistochastic and qustochastic matrices}

\author[ O. Chterental and D.\v{Z}. \DJ okovi\'{c}]
{Oleg Chterental and Dragomir \v{Z}. \DJ okovi\'{c}}

\address{Department of Pure Mathematics, University of Waterloo,
Waterloo, Ontario, N2L 3G1, Canada}

\email{ochteren@uwaterloo.ca} \email{djokovic@uwaterloo.ca}

\thanks{
The first author was supported by an NSERC Undergraduate Student Research Award, and the
second by NSERC Discovery Grant A-5285}


\date{}

\begin{abstract}
We introduce qustochastic matrices as the bi\-sto\-cha\-stic matrices arising from quaternionic unitary matrices by repla\-cing each entry with the square of its norm. This is the qua\-ter\-ni\-onic analogue of the unistochastic matrices studied by physicists. We also introduce qua\-ter\-ni\-onic Hadamard matrices and quaternionic mutually unbiased bases (MUB). In particular we show that the number of MUB in an $n$-dimensional quaternionic Hilbert space is at most $2n+1$. The bound is attained for $n=2$. We also de\-ter\-mine all quaternionic Hadamard matrices of size $n \leq 4$.
\end{abstract}
\maketitle \subjclassname { 81P68, 52B11}

\section{Introduction}\label{intro}

Let $\pB_n$ be the set of $n$ by $n$ bistochastic (also known as doubly stochastic) matrices, i.e., real matrices all of whose entries are nonnegative and row and column sums are identically $1$. The set $\pB_n$ is also known as the Birkhoff polytope; it is exactly the convex hull of the $n$ by $n$ permutation matrices. We shall use the standard notation $\Ort(n)$, $\U(n)$ and $\Sp(n)$ for the classical compact Lie groups: the real orthogonal, complex unitary and the quaternionic unitary (symplectic) group respectively. Let $\phi_r: \Ort(n) \rightarrow \pA_n$, $\phi_c: \U(n) \rightarrow \pA_n$ and $\phi_h: \Sp(n) \rightarrow \pA_n$ be the maps given by $\phi(Z)_{i,j}=|Z_{i,j}|^2$ where $\pA_n$ is the affine space spanned by $\pB_n$. The images of the first two maps are called the {\em orthostochastic} and {\em unistochastic} matrices respectively. We introduce the term {\em qustochastic} for matrices belonging to the image of $\phi_h$. We warn the reader that the term ``orthostochastic" has been used in the past by mathematicians, e.g. in \cite{AYC}, to designate the class of bistochastic matrices to which we refer now as ``unistochastic".

We will use the notation $\pO_n$, $\pU_n$ and $\pQ_n$ for the $n$ by $n$ orthostochastic, unistochastic and qustochastic matrices, respectively. Clearly we have $\pO_n \subseteq \pU_n \subseteq \pQ_n \subseteq \pB_n$. If $n=2$ all of these are equal. It is known that $\pO_n \subset \pU_n \subset \pB_n$ for $n \geq 3$. It is easy to see that $\pU_3 = \pQ_3$ and we will show that $\pU_n \subset \pQ_n \subset \pB_n$ for $n > 3$.

To {\em dephase} a complex matrix means to multiply it on the left and right by unitary diagonal matrices so that the uppermost row and the leftmost column become both real and nonnegative. We say that two matrices are {\em equivalent} if one can be obtained from the other by multiplication on the left and right by unitary monomial matrices. Similar definitions apply to the real and quaternion cases using real orthogonal and symplectic monomial matrices respectively. Note that in the quaternion case these operations allow the conjugation of all entries by a fixed unit norm quaternion. A class of equivalent matrices under the above relation may contain more than one dephased form. For example the dephased form may change under a permutation of rows and columns. If $a$ is a quaternion then we write it in coordinates with respect to the standard basis as $a=a_1 + a_2{\bf i} + a_3{\bf j} + a_4{\bf k}$. We use the notation $a^{\rm pu}$ for the pure part of $a$.

In section \ref{crit points} we determine the singular points of the map $\phi_r$ for $n \leq 4$ and the critical points of $\phi_c$ and $\phi_h$ for $n \leq 3$. These critical points are important if one wants to determine the boundaries of $\pU_n$ and $\pO_n$ in $\pA_n$. Indeed, if $Z \in \U(n)$ and $\phi_c(Z) \in \partial \pU_n$, the boundary of $\pU_n$, then $Z$ is necessarily a critical point of $\phi_c$.

The van der Waerden matrix $J_n \in \pB_n$, all of whose entries are equal to $1/n$, is the barycenter of $\pB_n$. We have that $J_n \in \pU_n$ for all $n$ but $J_n \in \pO_n$ if and only if there exists a real Hadamard matrix of order $n$. As $J_3 \not \in \pO_3$ it is an interesting question to find the distance from $J_3$ to $\pO_3$ (see Proposition \ref{distJ}). We also consider criteria for certain bistochastic matrices to be orthostochastic. It is easy to construct a system $\Sigma_n$ of $n(n-1)$ polynomial equations that must be satisfied by an $n$ by $n$ orthostochastic matrix. We determine that for $n \geq 16$ this system is not sufficient to characterize the set of orthostochastic matrices. This is done in section \ref{ortho matr}.

The matrices $A = \sqrt n Z$ where $Z \in \U(n)$ and $\phi_c (Z) = J_n$ are the complex Hadamard matrices. They satisfy $A^*A=nI_n$ and $|A_{ij}|=1$ for all $i,j$. In section \ref{quat hadamard} we define similarly the quaternionic Hadamard matrices. Thus, up to the scale factor $\sqrt n$, the set of complex Hadamard matrices of order $n$ can be identified with the fibre of $\phi_c$ above the point $J_n$.

While the real Hadamard matrices have been studied for more than $100$ years, and the complex ones for several decades, the quaternionic ones have been totally neglected. Every quaternionic Hadamard matrix is equivalent to a complex one if $n \leq 3$, but not so for $n \geq 4$. In the case $n=4$ we have constructed two families of quaternionic Hadamard matrices: a generic one depending on three real parameters and a special one having only two parameters. Any quaternionic Hadamard matrix of order $4$ is equivalent to a matrix belonging to one of these two families.

In section \ref{quat mub} we extend the concept of mutually unbiased bases (MUB) from the case of an $n$-dimensional complex Hilbert space to the quaternionic case. It is well known that in the complex case the number of MUB of $\bC^n$ does not exceed $n+1$ and that this bound is attained when $n$ is a power of a prime number. In the quaternionic case this bound is no longer valid and we obtain the new bound $2n+1$ i.e., for the quaternionic Hilbert space $\bH^n$. This new bound is attained for $n=2$, i.e., for $2$ by $2$ quaternionic matrices. The case $n=3$ is much harder.

In order to construct MUB $(A_0, A_1, \ldots, A_k)$ consisting of $3$ by $3$ symplectic matrices, we may assume without any loss of generality that $A_0 = I_3$, and $A_1 = F_3$ is the complex Fourier matrix. We have classified all possible members $A_2$ in such MUB by constructing a generic $3$-parameter family and five $2$-parameter families. Up to conjugation by a fixed complex number, $\sqrt 3 A_2$ is equivalent to a matrix in one of these families. Using this we were able to verify that a set of complex MUB consisting of four bases (there is only one such set up to equivalence) is maximal i.e. cannot be extended to quaternionic MUB of larger size. We also find a $3$-parameter set of four quaternionic MUB and show that they are all maximal.

The more complicated calculations and proofs were verified with Maple \cite{M}. Throughout the paper we occasionally remind the reader that a calculation was carried out on the computer.

\section{Critical Points}\label{crit points}

We will examine the critical points of the maps $\phi_c$ and $\phi_h$ (and singular points of $\phi_r$) in small dimensions $n$. The definition of singular points will be given later. For the tangent space ${\rm T}_B\pA_n$ of $\pA_n$ at $B \in \pA_n$ we use the basis of matrices $B^{i,j}$ where \[B^{i,j}_{k,l}=\left\{\begin{array}{ll} 1 & {\rm if \ (k,l)=(i,j) \ or \ (n,n)} \\ -1 & {\rm if \ (k,l)=(i,n) \ or \ (n,j)} \\ 0 & {\rm else} \end{array} \right.\] and $1 \leq i,j \leq n-1$.

To describe the various tangent spaces, let us introduce the following matrices $A^{i,j}$ and $C^{i,j}$ by specifying their entries: \[ \begin{array}{cc} A^{i,j}_{k,l}=\left\{\begin{array}{ll} 1 & {\rm if \ (k,l)=(i,j)} \\ -1 & {\rm if \ (k,l)=(j,i)} \\ 0 & {\rm else} \end{array} \right., & C^{i,j}_{k,l}=\left\{\begin{array}{ll} 1 & {\rm if \ (k,l)=(i,j) \ or \ (j,i)} \\ 0 & {\rm else} \end{array} \right. \\ 1 \leq i < j \leq n & 1 \leq i \leq j \leq n. \end{array}\] 
A basis of the tangent space ${\rm T}_X\Ort(n)$ at $X \in \Ort(n)$ is given by \[A^{i,j}X, \quad i<j.\] A basis of the tangent space ${\rm T}_Z\U(n)$ at $Z \in \U(n)$ is given by \[A^{i,j}Z, \quad i<j; \quad {\bf i}C^{i,j}Z, \quad i \leq j.\] Similarly a basis for the tangent space ${\rm T}_W\Sp(n)$ at $W \in \Sp(n)$ is given by \[A^{i,j}W, \quad i<j; \quad {\bf i}C^{i,j}W, \quad {\bf j}C^{i,j}W, \quad {\bf k}C^{i,j}W, \quad i \leq j.\] Now $\phi_r$ is obtained by restricting the codomain of the composition $f_r \circ i$ where $f_r : M_n(\bR) \rightarrow M_n(\bR)$ squares each entry and $i:\Ort(n) \rightarrow M_n(\bR)$ is the inclusion map. So we can find the differential $({\rm d}f_r)_X$ of $f$ at $X$ and then restrict to the tangent spaces to get the differential $({\rm d}\phi_r)_X:{\rm T}_X\Ort(n) \rightarrow {\rm T}_{\phi_r(X)}\pA_n$.

With a suitable ordering of the basis vectors, the differential $({\rm d}f_r)_X$ is represented by the $n^2$ by $n^2$ diagonal matrix $2 \cdot {\rm diag}(x_{1,1},x_{1,2},...,x_{n,n})$. In fact we have \begin{equation}\label{rdiff} ({\rm d}f_r)_X(Y)=X \circ Y, \end{equation} where $X \circ Y$ denotes the Hadamard product of matrices $X$ and $Y$. Explicitly, we have that \[ X \circ (A^{i,j}X) = \begin{bmatrix} 0 & 0 & \cdots & 0 \\ \vdots & \vdots & & \vdots \\ x_{i,1}x_{j,1} & x_{i,2}x_{j,2} & \cdots & x_{i,n}x_{j,n} \\ \vdots & \vdots & & \vdots \\ -x_{j,1}x_{i,1} & -x_{j,2}x_{i,2} & \cdots & -x_{j,n}x_{i,n} \\ \vdots & \vdots & & \vdots \\ 0 & 0 & \cdots & 0 \end{bmatrix}\] where entries are zero except in the $i^{th}$ and $j^{th}$ rows. Then it is not difficult to see that \[X \circ (A^{i,j}X) = \sum_{k=1}^{n-1} x_{i,k}x_{j,k}(B^{i,k} - B^{j,k})\] when $j<n$ and \[X \circ (A^{i,n}X) = \sum_{k=1}^{n-1} B^{i,k}x_{i,k}x_{n,k}. \] These expressions will be used to give the matrix representation of $({\rm d}\phi_r)_X$ with respect to the described bases.

In the complex case, the Jacobian of the map $\bC \rightarrow \bR$ given by $z_{ij}=x_{ij}+y_{ij}{\bf i} \mapsto |z_{ij}|^2=x_{ij}^2+y_{ij}^2$ is the matrix \[ \begin{bmatrix} 2x_{ij} \\ 2y_{ij} \end{bmatrix}. \] It sends the tangent vector $w_{ij} = u_{ij}+v_{ij}{\bf i}$ to \[ \begin{bmatrix} u_{ij} & v_{ij} \end{bmatrix} \begin{bmatrix} 2x_{ij} \\ 2y_{ij} \end{bmatrix} = 2 {\rm Re}(\bar z_{ij} w_{ij}). \] It follows that \[ ({\rm d}f_c)_Z(X) = {\rm Re}(\bar Z \circ X) \] and consequently \begin{equation}\label{basiseq}({\rm d} \phi_c)_Z(XZ) = {\rm Re} (\bar Z \circ (XZ)) \end{equation} for any skew-hermitian matrix $X$ and $Z \in \U(n)$. If $i=j$ we see that \[ ({\rm d} \phi_c)_Z({\bf i}C^{i,j}Z)=0. \] If $W$ is symplectic we mention that the analog of (\ref{basiseq}) holds.

Note that if one is to consider the Jacobian of $\phi_c$ or $\phi_h$ at a real orthogonal matrix, it is the Jacobian of $\phi_r$ with extra columns of zeros appended and similarly for unitary matrices under $\phi_h$. Thus if $X \in \Ort(n)$ then \[ {\rm rank}({\rm d} \phi_h )_X = {\rm rank}({\rm d} \phi_c )_X = {\rm rank}({\rm d} \phi_r )_X, \] and if $Z \in \U(n)$ then \begin{equation}\label{rank} {\rm rank}({\rm d} \phi_h )_Z = {\rm rank}({\rm d} \phi_c )_Z . \end{equation}

We say that $Z \in \U(n)$ is a {\em critical point} of $\phi_c$ if ${\rm rank}({\rm d} \phi_c)_Z < \dim \pA_n = (n-1)^2$. We define similarly the critical points of $\phi_h$. We say that $X \in \Ort(n)$ is a {\em singular point} of $\phi_r$ if ${\rm rank}({\rm d} \phi_r)_X < \dim \Ort(n) = n(n-1)/2$. Note that strictly speaking, a critical point of $\phi_r$ is a point $X \in \Ort(n)$ such that ${\rm rank}({\rm d \phi_r})_X < \dim \pA_n = (n-1)^2$. However since $\dim \Ort(n) < \dim \pA_n$ for $n \geq 3$, we would have that all points are critical for $n \geq 3$. Thus for $\phi_c$ and $\phi_h$ a critical point is one where the differential is not surjective, while for $\phi_r$ we use the term singular point when the differential is not injective.

\begin{remark}
From formula (\ref{rdiff}) we see that $({\rm d} f_r)_X$ is an isomorphism if and only if $X$ has no zero entries. In particular if $X \in \Ort(n)$ is a singular point of $\phi_r$, then $X$ must have a zero entry. Hence the set of nonsingular points of $\phi_r$ is an open dense subset of $\Ort(n)$.

The analogous question for $\phi_c$ (and $\phi_h$) remains open, see Problem \ref{sing}.
\end{remark}

We say that a square matrix $A$ {\em splits} if there exist permutation matrices $P$ and $Q$ such that $PAQ$ is a direct sum of two square matrices of smaller size. The set of critical points is invariant under equivalence \cite[3.1 and 3.2]{TZS}.  We have the following fact:

\begin{theorem}\label{split}
If an orthogonal matrix splits, then it is a singular point of the map $\phi_r$. Similar statements hold for the critical points of the maps $\phi_c$ and $\phi_h$.
\end{theorem}
\begin{proof}
Let $X \in \Ort(n)$ split and since $X$ is a singular point iff $PXQ$ is a singular point for permutation matrices $P$ and $Q$, we assume that $X$ is a direct sum of square matrices of smaller size. Assume $1 \leq j < n$ is the smallest index such that $X$ is the direct sum of a $j$ by $j$ block and an $n-j$ by $n-j$ block. Then $X_{k,l}=0$ for $k>j$, $l \leq j$ and for $k \leq j$, $l > j$. Consider the basis vector $A^{1,j+1}X$ which has first row equal to the $(j+1)^{st}$ row of $X$ and has $(j+1)^{st}$ row equal to the negative of the first row of $X$ and all other rows are zero. Now it is not difficult to see that the Hadamard product $X \circ (A^{1,j+1}X)$ is the zero matrix. Thus $A^{1,j+1}X \in {\rm ker}({\rm d}\phi_r)_X$ and $X$ is a singular point of the map $\phi_r$.

If $Z \in \U(n)$ splits we must show that ${\rm rank}({\rm d}\phi_c)_Z < (n-1)^2$. Again assume $1 \leq j < n$ is the smallest index such that $Z$ is the direct sum of a $j$ by $j$ block and an $n-j$ by $n-j$ block. Then $Z_{k,l}=0$ and using (\ref{basiseq}) we see that $A^{k,l} \in {\rm ker}({\rm d}\phi_c)_Z$ for $k>j$, $l \leq j$ and for $k \leq j$, $l > j$. Also $({\rm d} \phi_c )_Z({\bf i}C^{i,i}Z)=0$ for $1 \leq i \leq n$. Thus ${\rm dim \ ker}({\rm d}\phi_c)_Z \geq 2(n-j)j+n \geq 2n$ and $Z$ is a critical point.

If $W \in \Sp(n)$ then the proof is similar to the complex case using the analog of (\ref{basiseq}).
\end{proof}

Now we will find the singular (critical) points of $\phi_r$ ($\phi_c$ and $\phi_h$) for $n \leq 3$. For a real or complex matrix $X$, we denote by \[X \begin{pmatrix} i_1, \dots, i_k \\ j_1,\dots,j_k \end{pmatrix}\] its $k$ by $k$ minor in rows $i_1, \dots, i_k$ and columns $j_1, \dots, j_k$.

\begin{theorem}\label{ortho uni cp n=2}
Let $n=2$. The singular (critical) points of $\phi_r$ ($\phi_c$ and $\phi_h$) are the diagonal and anti-diagonal matrices.
\end{theorem}
\begin{proof}
Let
\[ X=\begin{bmatrix} 
x_{1,1} & x_{1,2} \\
x_{2,1} & x_{2,2}
\end{bmatrix} \]
be a real orthogonal matrix. The Jacobian of $\phi_r$ at $X$ is the $1$ by $1$ matrix $[x_{1,1}x_{2,1}]$. It is clear then that there are $8$ critical points given by the matrices
\[ \begin{bmatrix} 
\pm 1 & 0 \\
0 & \pm 1
\end{bmatrix}, \quad
\begin{bmatrix} 
0 & \pm 1 \\
\pm 1 & 0
\end{bmatrix}. \] It is not difficult to see that this reasoning carries over to the complex and quaternion cases since any $2$ by $2$ unitary or symplectic matrix is equivalent to a real orthogonal matrix.
\end{proof}

\begin{theorem}\label{ortho cp n=3}
Let $n=3$. The singular points of the map $\phi_r$ are the matrices that split.
\end{theorem}
\begin{proof}
It is clear that any matrix that splits is a singular point. Let $X=[x_{i,j}] \in \Ort(3)$ be a singular point. The Jacobian of $\phi_r$ at $X$ is the $4$ by $6$ matrix (three columns of zeros ignored):
\[ J=\begin{bmatrix}
x_{1,1}x_{2,1} & x_{1,1}x_{3,1} & 0 \\
x_{1,2}x_{2,2} & x_{1,2}x_{3,2} & 0 \\
-x_{1,1}x_{2,1} & 0 & x_{2,1}x_{3,1} \\
-x_{1,2}x_{2,2} & 0 & x_{2,2}x_{3,2}
\end{bmatrix} \]
If $X$ has two or more zeros it must split. Assume $X$ has at most one zero which we may move to the third column. Permute the rows of $X$ so that \[ X \begin{pmatrix} 1,3 \\ 1,2 \end{pmatrix} \neq 0. \] Then \[ J \begin{pmatrix} 2,3,4 \\ 1,2,3 \end{pmatrix} = x_{1,2}x_{3,2}x_{2,1}x_{2,2} X \begin{pmatrix} 1,3 \\ 1,2 \end{pmatrix} \] so \[ x_{1,2}x_{3,2}x_{2,1}x_{2,2} = 0, \] a contradiction.
\end{proof}

\begin{theorem}\label{unitary symplectic cp n=3}
Let $n=3$. The set of critical points for the maps $\phi_c$ and $\phi_h$ is, up to equivalence, the set $\Ort(3)$ of $3$ by $3$ real orthogonal matrices.
\end{theorem}
\begin{proof}
We consider the complex case first. Note that each $Z \in \Ort(3)$ is a critical point of $\phi_c$ (see \cite[Lemma 3.7]{TZS}) and as mentioned earlier, each $X \in \U(3)$ which is equivalent to $Z$ is also a critical point of $\phi_c$.

To prove the converse, consider an arbitrary critical point in dephased form:
\[ Z=\begin{bmatrix} 
x_{1,1} & x_{1,2} & x_{1,3} \\
x_{2,1} & x_{2,2}+y_{2,2}\bf{i} & x_{2,3}+y_{2,3}\bf{i} \\
x_{3,1} & x_{3,2}+y_{3,2}\bf{i} & x_{3,3}+y_{3,3}\bf{i} \\
\end{bmatrix}. \]
Construct the $4$ by $9$ Jacobian (note that three columns are zero and can be ignored):
\[ J=\begin{bmatrix} 
x_{1,1}x_{2,1} & x_{1,1}x_{3,1} & 0 & 0 & 0 & 0 \\
x_{1,2}x_{2,2} & x_{1,2}x_{3,2} & 0 & -x_{1,2}y_{2,2} & -x_{1,2}y_{3,2} & 0 \\
-x_{1,1}x_{2,1} & 0 & x_{2,1}x_{3,1} & 0 & 0 & 0 \\
-x_{1,2}x_{2,2} & 0 & \alpha & x_{1,2}y_{2,2} & 0 & \beta
\end{bmatrix} \]
where $\alpha=x_{2,2}x_{3,2}+y_{2,2}y_{3,2}$ and $\beta=y_{2,2}x_{3,2}-x_{2,2}y_{3,2}$. Then all $4$ by $4$ minors of $J$ are $0$. In particular \[J \begin{pmatrix} 1,2,3,4 \\ 1,2,4,5 \end{pmatrix} = x_{1,1}^2x_{2,1}x_{3,1}x_{1,2}^2y_{2,2}y_{3,2}=0. \]

We have two cases.

Case $1$: $Z$ has a row with all entries not equal to $0$. Then since $Z$ is unitary it can be seen that it must have a column with all entries not equal to $0$. After permuting rows and columns (independently) and dephasing, we may assume that all terms $x_{i,1},x_{1,i}>0$. According to the above minor, we see that at least one of $y_{2,2}$ or $y_{3,2}$ must be $0$. Then using the fact that the rows and columns of $Z$ are orthogonal we conclude that the remaining $y_{i,j}$ must also be $0$ and $Z$ is in fact real and orthogonal.

Case $2$: $Z$ has a zero in each row. Then since $Z$ is unitary we see it must split.

Now we consider the quaternionic case. By dephasing a $3$ by $3$ symplectic matrix $W$ and then conjugating each entry by a fixed quaternion of unit norm so as to make the $(2,2)$ entry complex, we see that $W$ is equivalent to a complex matrix $Z$. Thus if $W$ is a critical point of $\phi_h$ so is $Z$. By (\ref{rank}) we have that $Z$ is also a critical point of $\phi_c$ and thus is equivalent to a real matrix.
\end{proof}

We now take a look at the singular points of $\phi_r:\Ort(4) \rightarrow \pA_4$. 

\begin{theorem}\label{ortho cp n=4}
Let $n=4$. There are two types of singular points of the map $\phi_r$. The first are the matrices that split. The second type are the matrices which are equivalent to a matrix having all diagonal entries zero and all other entries nonzero.
\end{theorem}
\begin{proof}
By theorem \ref{split}, all orthogonal matrices that split are singular points. Also it is clear after a computation that matrices of the second type are also singular points. Assume $X=[x_{i,j}]\in \Ort(4)$ is a singular point. The Jacobian of $\phi_r$ at $X$ is (ignoring zero columns):
\[ J=\begin{bmatrix} 
x_{1,1}x_{2,1} & x_{1,1}x_{3,1} & x_{1,1}x_{4,1} & 0 & 0 & 0 \\
x_{1,2}x_{2,2} & x_{1,2}x_{3,2} & x_{1,2}x_{4,2} & 0 & 0 & 0 \\
x_{1,3}x_{2,3} & x_{1,3}x_{3,3} & x_{1,3}x_{4,3} & 0 & 0 & 0 \\
-x_{2,1}x_{1,1} & 0 & 0 & x_{2,1}x_{3,1} & x_{2,1}x_{4,1} & 0 \\
-x_{2,2}x_{1,2} & 0 & 0 & x_{2,2}x_{3,2} & x_{2,2}x_{4,2} & 0 \\
-x_{2,3}x_{1,3} & 0 & 0 & x_{2,3}x_{3,3} & x_{2,3}x_{4,3} & 0 \\
0 & -x_{3,1}x_{1,1} & 0 & -x_{3,1}x_{2,1} & 0 & x_{3,1}x_{4,1} \\
0 & -x_{3,2}x_{1,2} & 0 & -x_{3,2}x_{2,2} & 0 & x_{3,2}x_{4,2} \\
0 & -x_{3,3}x_{1,3} & 0 & -x_{3,3}x_{2,3} & 0 & x_{3,3}x_{4,3} \\
\end{bmatrix}. \]
All $6$ by $6$ minors of $J$ vanish. By evaluating \[ J \begin{pmatrix} 1,2,3,4,5,k \\ 1,2,3,4,5,6 \end{pmatrix}; \quad k=7,8,9, \] we obtain \begin{equation}\label{minoreq}x_{1,1}x_{1,2}x_{1,3}x_{2,1}x_{2,2}x_{3,k}x_{4,k} X \begin{pmatrix} 3,4 \\ 1,2 \end{pmatrix} X \begin{pmatrix} 2,3,4 \\ 1,2,3 \end{pmatrix} = 0. \end{equation} Notice that the last two factors are nested minors of $X$. Whenever it is possible to find two nested nonzero minors of $X$ we may permute rows and columns to assume that they are \[ X \begin{pmatrix} 3,4 \\ 1,2 \end{pmatrix} \quad {\rm and} \quad X \begin{pmatrix} 2,3,4 \\ 1,2,3 \end{pmatrix},\] and then use the equations (\ref{minoreq}) to conclude the existence of zero entries in $X$. Now we consider cases depending on how many zero entries are in $X$. If a $4$ by $4$ real orthogonal matrix has more than $4$ zero entries, it splits.

Case $1$: $X$ has at most one zero entry. We may assume the zero entry is in the fourth column if it exists. Since $X$ is orthogonal, we may find the required nested nonzero $3$ by $3$ and $2$ by $2$ minors amongst the first three columns of $X$. Then permuting the rows of $X$ and the first three columns as necessary we may assume that the nested nonzero minors are in the top left corner of $X$. Then (\ref{minoreq}) gives a contradiction.

Case $2$: $X$ has exactly $2$ zero entries. If they are in the same row/column we may permute and take transpose as necessary to have them in the $(3,4)$ and $(4,4)$ positions and repeat the above argument to get a contradiction. Otherwise we place them in the $(3,3)$ and $(4,4)$ positions. Let the four rows of the first three columns be $a$, $b$, $c$ and $d$. We now must find a nonzero $3$ by $3$ minor of the first three columns containing row $c$. Assume all such minors are zero. Consider the minors using $\{a,b,c\}$ and $\{b,c,d\}$. We have that \[ c = \alpha a +\beta b = \gamma b + \delta d\] for some $\alpha, \beta, \gamma, \delta \in \bR$. We get \[ \alpha a + (\beta - \gamma)b - \delta d =0,\] but since $X$ is nonsingular, the $\{a,b,d\}$ minor is nonzero and so $\alpha = 0$. Thus $c = \beta b$ but clearly this is a contradiction since $c$ has a zero entry. Now we may also find a nonzero $2$ by $2$ nested minor in the first two columns. Then equations (\ref{minoreq}) give a contradiction.

Case $3$: $X$ has exactly $3$ zero entries. Clearly they are not all in one row or column. If two are in the same row/column we may assume they are in the fourth column and we use the arguments from the previous case to get a contradiction. Thus assume the zeros appear in $(2,2)$, $(3,3)$ and $(4,4)$. Let the four rows of the first three columns be $a$, $b$, $c$ and $d$. We now must find a nonzero $3$ by $3$ minor of the first three columns containing rows $b$ and $c$. Assume all such minors are zero. Then both $a$ and $d$ would be in ${\rm span}\{b,c\}$, which is clearly impossible. As the nested $2$ by $2$ minor contained in rows $b$ and $c$ is nonzero, the equations (\ref{minoreq}) give a contradiction.

Case $4$: By now we see that we may assume that the $4$ zeros are on the main diagonal and we are done.
\end{proof}

\begin{remark}
If $X \in \Ort(n)$ has zero diagonal, then all principal $n-1$ by $n-1$ minors of $X$ vanish. This follows from Cramer's formula for the entries of $X^{-1}$ and the fact that $X^{-1}=X^T$ also has zero diagonal.
\end{remark}

\section{Orthostochastic Matrices}\label{ortho matr}

Consider an aribtrary $3$ by $3$ bistochastic matrix:
\begin{equation}\label{bistoch} \begin{bmatrix}
x & y & \star  \\
z & w & \star \\
a & b & \star \\
\end{bmatrix}.\end{equation}

It is orthostochastic if and only if the vectors \[ (\sqrt x, \sqrt z, \sqrt a ) \] and \[ (\sqrt y, \pm \sqrt w, \pm \sqrt b) \] are orthogonal for some appropriate choice of signs. That is \[ \sqrt{xy} + \varepsilon_1 \sqrt{zw} + \varepsilon_2 \sqrt{ab} = 0 \] for some $\varepsilon_i=\pm 1$. Squaring gives \[ xy + zw + 2\varepsilon_1 \sqrt{xyzw} = ab. \] Substituting $a=1-x-z$ and $b=1-y-w$ and squaring again gives \begin{equation}\label{orthoeq}(1-x-y-z-w+xw+yz)^2=4xyzw.\end{equation}

We are thus lead to the following result of H. Nakazato \cite{N}. Apparently he failed to notice that the defining equation of $\pO_3$ can be written in this simple form. A similar expression appears in \cite{D} as an inequality describing $\pU_3$.

\begin{proposition}
A $3$ by $3$ bistochastic matrix (\ref{bistoch}) is orthostochastic if and only if it satisfies the equation (\ref{orthoeq}).
\end{proposition}

Since $J_3 \not \in \pO_3$ we find the distance between $J_3$ and the set $\pO_3$.

\begin{proposition}\label{distJ}
The distance from $J_3$ to $\pO_3$ is $\sqrt 2 / 3$ and the closest points on $\pO_3$ are
\begin{equation}\label{closest} \frac{1}{9}
\begin{bmatrix}
1 & 4 & 4 \\
4 & 1 & 4 \\
4 & 4 & 1 \\
\end{bmatrix}
\end{equation} and all points obtained by permuting the rows and columns of this matrix.
\end{proposition}
\begin{proof}
Let $B$ be a bistochastic matrix as in (\ref{bistoch}). We look for the minimum of the squared distance \[ d(B,J_3)^2 = f(x,y,z,w) \] under the constraint (\ref{orthoeq}). Using the method of Lagrange multipliers we get a system of $5$ equations
which Maple is able to solve. Up to permutations of the rows and columns, there are two stationary points. One is \[ \frac{1}{4}
\begin{bmatrix}
1 & 2 & 1 \\
2 & 0 & 2 \\
1 & 2 & 1 \\
\end{bmatrix}
\] at a distance of $\frac{1}{2}$ from $J_3$ and the other is the matrix (\ref{closest}) at a distance of $\sqrt 2 / 3$ from $J_3$. By using the Hessian test for optimization problems with a constraint, one can check that the first of these points is a saddle point and the second a local minimum. Note that the first point is the barycenter of one of the faces of $\pB_3$ and so it is the closest point on $\partial \pB_3$ to $J_3$. We can now conclude that $f$ has absolute minimum at the second point.
\end{proof}

This result is stated in \cite{BEKTZ}. It is mentioned to have been deduced from a similar problem involving minimizing entropy, which was solved numerically in \cite{GMPS}.

Now consider an arbitrary $4$ by $4$ bistochastic matrix:
\begin{equation}\label{bistoch2} \begin{bmatrix}
x & u & p & \star \\
y & v & q & \star \\
z & w & r & \star \\
a & b & c & \star \\
\end{bmatrix}. \end{equation}
It is orthostochastic if and only if the vectors \[ (\sqrt x, \sqrt y, \sqrt z, \sqrt a), \] \[ (\sqrt u, \varepsilon_1 \sqrt v, \varepsilon_2 \sqrt w, \varepsilon_3 \sqrt b), \] and \[ (\sqrt p, \nu_1 \sqrt q, \nu_2 \sqrt r, \nu_3 \sqrt c) \] are orthogonal for some $\varepsilon_i, \nu_i = \pm 1$. The existence of $\varepsilon_i$ to make the first two vectors orthogonal is given by the equation \[ \sqrt{xu} + \varepsilon_1 \sqrt{yv} + \varepsilon_2 \sqrt{zw} + \varepsilon_3 \sqrt{ab} = 0 \] which is equivalent to \[ xu+yv+2 \varepsilon_1 \sqrt{xuyv} = zw+ab+ 2 \varepsilon_2 \varepsilon_3 \sqrt{zwab}. \] Then rearranging and squaring again gives \[ (xu+yv-zw-ab)^2 = 4xuyv + 4zwab - 2 \varepsilon_1 \varepsilon_2 \varepsilon_3 \sqrt{xuyvzwab}. \] Finally squaring one last time and substituting $a=1-x-y-z$ and $b=1-u-v-w$ gives \begin{equation}\label{4bist} [(xu+yv-zw-(1-x-y-z)(1-u-v-w))^2 \end{equation} \[ -4xuyv-4zw(1-x-y-z)(1-u-v-w)]^2 \] \[ = 4xuyvzw(1-x-y-z)(1-u-v-w). \] The expressions for the other pairs of vectors are similarly obtained. For example the first and third columns can be made orthogonal if and only if the equation \[  [(xp+yq-zr-(1-x-y-z)(1-p-q-r))^2 \] \[ -4xpyq-4zr(1-x-y-z)(1-p-q-r)]^2 \] \[ = 4xpyqzr(1-x-y-z)(1-p-q-r), \] is satisfied. We obtain all together six equations by considering all pairs of columns. As well we obtain six more equations by considering pairs of rows.

\begin{proposition}
If the matrix (\ref{bistoch2}) is orthostochastic matrix then (\ref{4bist}) and the eleven remaining equations hold.
\end{proposition}

\begin{remark}
The problem of determining if these twelve equations are sufficient to ensure that a $4$ by $4$ bistochastic matrix is orthostochastic, is open.
\end{remark}

For an $n$ by $n$ bistochastic matrix $X$ one can easily write a similar system $\Sigma_n$ of $n(n-1)$ equations. Let $X'$ be the $n$ by $n$ matrix with $X'_{ij}=\sqrt {X_{ij}}$. An equation in $\Sigma_n$ vanishes if and only if the corresponding pair of columns or rows in $X'$ can be made orthogonal by appropriately choosing signs for the entries. Thus an $n$ by $n$ orthostochastic matrix will satisfy $\Sigma_n$. Then a natural question arises: Does $\Sigma_n$ characterize $\pO_n$?

It is easy to see that the answer is negative in general. Indeed if $n \equiv 2 \pmod 4$ then $J_n$ satisfies $\Sigma_n$. On the other hand, if also $n>2$, $J_n \not \in \pO_n$ because there is no $n$ by $n$ Hadamard matrix.

Moreover there are also counterexamples for all $n \geq 16$. First we prove a lemma.

\begin{lemma}\label{algindep}
For $n > 1$ there exist positive real numbers $a_1, \ldots, a_n$ such that $\sum a_i = 1$, and the numbers $1$ and $\sqrt{a_i a_j}$ for $1 \leq i < j \leq n$ are linearly independent over the rational numbers. 
\end{lemma}
\begin{proof}
Choose positive real numbers $\xi_i$, $1 \leq i \leq n-1$ which, are algebraically
independent over $\bQ$, the rational numbers, and are such that $\sum \xi_i^2 < 1$. Let \[ a_i = \xi_i^2, 1 \leq i \leq n-1;  \quad a_n = 1-\sum_{i=1}^{n-1} \xi_i^2. \] 

To prove the assertion, assume that \[ r_0 + \sum _{1 \leq i < j \leq n} r_{i,j} \sqrt{a_i a_j} = 0 \] for some rational numbers $r_0$ and $r_{i,j}$. This can be rewritten as \[ r_0 + \sum_{1 \leq i < j \leq n-1} r_{i,j} \xi_i \xi_j = - \sum_{i=1}^{n-1} r_{i,n} \xi_i \sqrt{a_n}. \] Squaring both sides and rearranging gives \[ \left( r_0 + \sum_{1 \leq i < j \leq n-1} r_{i,j}\xi_i \xi_j \right)^2 -  \left( \sum_{i=1}^{n-1} r_{i,n} \xi_i \right)^2 \left( 1 - \sum_{i=1}^{n-1} \xi_i^2 \right) = 0. \] The coefficient of $\xi_i^2 \xi_j^2$ for $1 \leq i<j \leq n-1$ is $r_{i,j}^2+r_{i,n}^2$. Now since the $\xi_i$ are algebraically independent over $\bQ$ we have that $r_0$ and $r_{i,j}$ all vanish.
\end{proof}

Now we will construct a bistochastic matrix which satisfies $\Sigma_n$ but is not orthostochastic, using arguments from \cite{AYC}. For two matrices $A$, $B$ we denote by $\overline{AB}$ the convex hull of $\{A,B\}$.

\begin{proposition}
Let $n \geq 16$.
\begin{itemize}
\item[(i)] There exist $X$ in $\pB_n$ satisfying $\Sigma_n$ which are not in $\pO_n$.
\item[(ii)] There exist a set $S$ of permutations on $n$ letters such that $\overline{P_\sigma P_\tau}$ is contained in $\pO_n$ for any $\sigma,\tau$ in $S$, while the convex hull of $S$ is not contained in $\pO_n$.
\end{itemize}
\end{proposition}
\begin{proof}
To do this we will need some definitions and the Hurwitz-Radon theorem \cite{L}. Let $A$ be an elementary abelian subgroup of order $n=16$ of $S_n$. Choose some linear order $<$ on $A$ with the identity $\rm id$ being the smallest element and let $\sigma \rightarrow R_\sigma$ be the regular (permutation) representation of $A$. Let $\{a_\sigma\}_{\sigma \in A}$ be as in Lemma \ref{algindep} and let \[ X = \sum_{\sigma \in A} a_\sigma R_\sigma \in \pB_n. \] The rows and columns of $X$ are labelled based on the linear order from $A$. One can see that $X_{\alpha, \beta} = a_{\alpha \beta}$.

Consider distinct columns $\alpha$ and $\beta$ in $X$. For any $\sigma \in A$ there is a unique corresponding $\tau \in A$ such that $\sigma \alpha = \tau \beta$. Then $X_{\sigma,\alpha} = X_{\tau,\beta}$ and $X_{\tau,\alpha} = X_{\sigma,\beta}$ and it is easy to see that we may choose signs appropriately so that the corresponding columns of $X'$ are orthogonal. A similar argument applies to the rows. Thus $X$ satisfies $\Sigma_n$. Now choosing signs for all entries in $X'$ to produce an orthogonal matrix is equivalent to the existence of diagonal orthogonal matrices $D_\sigma$ such that \[ \sum_{\sigma \in A} \sqrt a_\sigma \tilde R_\sigma \in \Ort(n), \] where $\tilde R_\sigma = D_\sigma R_\sigma$. Assume such diagonal orthogonal matrices exist. We can clearly assume that $D_{\rm id} = I_n$ and $\tilde R_{\rm id} = R_{\rm id} = I_n$. Thus by orthogonality we have \[ \sum_{\sigma \in A} a_\sigma \tilde R_\sigma^T \tilde R_\sigma + \sum_{\sigma < \tau} \sqrt{ a_\sigma a_\tau } ( \tilde R_\sigma^T \tilde R_\tau + \tilde R_\tau^T \tilde R_\sigma) = I_n. \] The first summation simply equals $I_n$. Thus since $\sqrt{a_i a_j}$ are linearly independent over $\bQ$, we get \[ \tilde R_\sigma^T \tilde R_\tau + \tilde R_\tau^T \tilde R_\sigma = 0, \quad \sigma < \tau, \] and in particular \[ \tilde R_\sigma^T + \tilde R_\sigma = 0, \quad \sigma > \rm id, \] and \[ \tilde R_\sigma^T \tilde R_\sigma = - \tilde R_\sigma^2 = I_n, \quad \sigma > \rm id. \] Thus the Hurwitz-Radon equations are satisfied. However the Hurwitz-Radon number $\rho(16)=9$ is less than $15$. Since the number of matrices that can satisfy the Hurwitz-Radon equations is always less than or equal to $\rho(n)$, we have a contradiction.

The group $A$ immediately provides us with the set $S$ for the second part of the proposition. All elements of $A$ are commuting involutions and so by Theorem \ref{invol} each edge $\overline{P_\sigma P_\tau}$ for $\sigma,\tau \in A$ is orthostochastic, yet the matrix $X$ constructed above is not. 

We then extend the result to all larger dimensions $n > 16$ by exten\-ding each permutation $\sigma \in A$ so that $\sigma(i)=i$ for $i>16$.
\end{proof}

In \cite{AYC} it is asked if a set of permutation matrices containing only commuting involutions has a unistochastic convex hull. We have thus shown in the above proposition that the answer is false in the ortho\-stochastic case. Note that in that paper the term ``orthostochastic" corresponds to our term unistochastic.

\section{Quaternion Hadamard Matrices}\label{quat hadamard}

An $n$ by $n$ {\em quaternion Hadamard matrix} is a matrix $H \in M_n(\bH)$ such that each entry has unit norm $|H_{i,j}|=1$ and $H^*H=nI_n$ where $^*$ denotes conjugate transpose, i.e., $\frac{1}{\sqrt{n}}H$ is symplectic. The $n$ by $n$ Fourier matrix, $F_n$, is the unitary matrix given by \[[F_n]_{i,j}=\frac{1}{\sqrt n} \omega^{(i-1)(j-1)} \] where $\omega= e^{2\pi {\bf i} / n}$. Dephasing a $2$ by $2$ symplectic matrix gives a real matrix while dephasing and conjugating entrywise a $3$ by $3$ symplectic matrix yields a complex matrix. It follows that every $3$ by $3$ quaternion Hadamard matrix is equivalent to the Fourier matrix. Thus we will investigate the structure of $4$ by $4$ quaternion Hadamard matrices. 

Let us first introduce two families of such matrices. The $2$-parameter {\em special family} of $4$ by $4$ quaternion Hadamard matrices consists of matrices of the form
\[ \begin{bmatrix}
1 & 1 & 1 & 1 \\
1 & -1 & b & -b \\
1 & a & x & z \\
1 & -a & y & w
\end{bmatrix}, \]
where $a$ and $b$ are unit norm quaternions of the form $a = a_1 + a_2 {\bf i}$, $b = b_1 + b_2 {\bf j}$ and 
\begin{equation}\label{special}
\begin{array}{cc} \displaystyle x=-\frac{1}{2}(1+a+b-ab), \quad & \displaystyle z=-\frac{1}{2}(1+a-b+ab) \\ \cr \displaystyle y=-\frac{1}{2}(1-a+b+ab), \quad & \displaystyle w=-\frac{1}{2}(1-a-b-ab). \end{array}
\end{equation}

The $3$-parameter {\em generic family} of $4$ by $4$ quaternion Hadamard matrices consists of the matrices of the form
\[ \begin{bmatrix}
1 & 1 & 1 & 1 \\
1 & a & b & -1-a-b \\
1 & c & d & -1-c-d \\
1 & -1-a-c & -1-b-d & 1+a+b+c+d \\
\end{bmatrix} \]
where \[ b= \left( \frac{1 + \hat a}{|1 + \hat a|} {\bf i} \right)^2, \ c = \left( x\frac{1+a}{|1+a|} \right)^2, \ d = \left( x \frac{1 + \hat a}{|1 + \hat a|} {\bf i} \right)^2, \] $a \neq -1$ and $x$ are unit norm quaternions of the form $a = a_1 + a_2 {\bf i} + a_3 {\bf j}$ and $x = x_2 {\bf i} + x_3 {\bf j}$, and $\hat a = a_1 + a_2 {\bf i}$.

\begin{theorem}
The matrices in the special and generic families are Hadamard.
\end{theorem}
\begin{proof}
We verified using Maple that these families indeed consist of Hadamard matrices. It is also possible to do the verification by hand.
\end{proof}

In the case $n=4$ there are genuine $4$ by $4$ quaternion Hadamard matrices, and we see that this allows the set of qustochastic matrices to be larger than the unistochastic matrices.

\begin{theorem}
$\pU_n$ is a proper subset of $\pQ_n$ if $n \geq 4$.
\end{theorem}
\begin{proof}
Let $n=4$ and consider the matrix $H$ from the special family with $a={\bf i}$ and $\sqrt{2}b=1+{\bf j}$. Then the differential $({\rm d} \phi_h )_H$ is a $9$ by $36$ matrix which a computation shows to have rank $9$. To be specific, the $9$ by $9$ minor of $({\rm d} \phi_h )_H$ made up from columns corresponding to basis vectors $EH$ for $E \in \{ A^{1,4}, A^{2,3}, A^{2,4}, A^{3,4}, {\bf i}C^{1,4}, {\bf i}C^{2,4}, {\bf j}C^{1,4}, {\bf j}C^{2,4}, {\bf k}C^{3,4}\}$ is nonzero. Thus there is an open ball of qustochastic matrices in the Birkhoff polytope centered at the van der Waerden matrix $J_4$. This is not the case for unistochastic matrices for $n=4$ as seen in \cite{BEKTZ}. Thus $\pU_4$ is a proper subset of $\pQ_4$. The result easily extends to the cases $n>4$ by observing that the matrix $\phi_h(X) \oplus I_{n-4}$ is in $\pQ_n \setminus \pU_n$ if $X \in \Sp(4)$ is chosen so that $\phi_h(X) \not \in \pU_4$.
\end{proof}

\begin{example} To see that $\pQ_n$ is a proper subset of $\pB_n$ for $n \geq 3$ simply note that 
\[ \frac{1}{2}
\begin{bmatrix}
1 & 0 & 1 \\
1 & 1 & 0 \\
0 & 1 & 1 \\
\end{bmatrix},  \quad
\frac{1}{2} \begin{bmatrix}
1 & 0 & 0 & 1 \\
1 & 1 & 0 & 0 \\
0 & 1 & 1 & 0 \\
0 & 0 & 1 & 1 \\
\end{bmatrix}, \quad \ldots
\] are bistochastic, but obviously not qustochastic. This argument is the same as the one used in \cite[Theorem 5]{AYP}.
\end{example}

For a permutation $\sigma$ in the symmetric group on $n$ symbols, we let $P_\sigma$ be the $n$ by $n$ permutation matrix with entries $(P_\sigma)_{i,j} = \delta_{i,\sigma(j)}$. The map $\sigma \rightarrow P_\sigma$ is a group homomorphism.

\begin{theorem}\label{invol}
If $\overline {P_\sigma P_\tau} \subseteq \pQ_n$ then $\sigma^{-1}\tau$ is an involution. Conversely, if $\sigma^{-1}\tau$ is an involution, then $\overline {P_\sigma P_\tau} \subseteq \pO_n$.
\end{theorem}
\begin{proof}
Assume that $\overline{P_\sigma P_\tau} \subseteq \pQ_n$. It is easy to see that the matrix $A=\frac{1}{2}(P_\sigma + P_\tau)$ is, up to permutation of the rows and columns, a direct sum of blocks
\[ \begin{bmatrix}1\end{bmatrix}, \quad \frac{1}{2} \begin{bmatrix} 1 & 1 \\ 1 & 1 \end{bmatrix}, \quad
\frac{1}{2} \begin{bmatrix}
1 & 1 & 0 \\
1 & 0 & 1 \\
0 & 1 & 1 
\end{bmatrix}, \quad \ldots
\] of various sizes. Since $A \in \pQ_n$, only blocks of size $1$ or $2$ occur. This implies that $\sigma^{-1}\tau$ is an involution. Conversely if $\sigma^{-1}\tau$ is an involution then any matrix $p P_\sigma + (1-p) P_\tau$ for $0 < p < 1$ is (up to permutation of rows and columns) a direct sum of blocks 
\[ \begin{bmatrix}1\end{bmatrix} \quad {\rm or} \quad \begin{bmatrix} p & 1-p \\ 1-p & p \end{bmatrix} \] and is easily seen to be in $\pO_n$.
\end{proof}

These arguments are similar to those found in \cite{AYC}.

Now we will show that the special and generic families comprise all $4$ by $4$ quaternion Hadamard matrices up to equivalence. We begin with a lemma:

\begin{lemma}\label{column form II}
Let ${\bf u}=[1,a,x,y]^T$ be a $4$ by $1$ quaternionic column vector with entries of unit norm. Assume ${\bf u}$ is orthogonal to the column vector with all entries equal to $1$ and $a \neq 1$. Then \begin{equation} x=h-\frac{1+a}{2}, \quad y=-h-\frac{1+a}{2}, \end{equation} where h is orthogonal to $1+a$ and $2|h|=|1-a|$.
\end{lemma}
\begin{proof}
Let $h=x+\frac{1+a}{2}$. Note that $1+a=-x-y$, $2h=x-y$ and $|x|=|y|=1$ so $h \perp 1+a$. Then $|h-\frac{1+a}{2}|^2=|x|^2=1$ so $|h|^2=1-|\frac{1+a}{2}|^2=|\frac{1-a}{2}|^2$ since $1+a \perp 1-a$. Thus $2|h|=|1-a|$. The expression for $y$ is straightforward.
\end{proof}

\begin{theorem}
Each $4$ by $4$ quaternion Hadamard matrix is equivalent to one in the special or generic family.
\end{theorem}
\begin{proof}
After dephasing a $4$ by $4$ quaternion Hadamard matrix we have one of the form
\[ \begin{bmatrix}
1 & 1 & 1 & 1 \\
1 & a & d & g \\
1 & b & e & h \\
1 & c & f & i
\end{bmatrix}. \]
If one of the nine entries $a,b,...,i$ is real, it must be $\pm 1$. When there is a real entry there must be a $-1$ entry also. Let us assume there is a real entry. After permuting rows and columns we obtain a matrix of the form
\[ \begin{bmatrix}
1 & 1 & 1 & 1 \\
1 & -1 & b & -b \\
1 & a & x & z \\
1 & -a & y & w
\end{bmatrix}. \]
We may assume that $a \in \bC$ and $b \in {\rm span}\{1,{\bf i},{\bf j}\}$ after conjugating. Using orthogonality with the first row and column, and solving for $x,y$ and $z$, we get \[x=-b-a+w, \quad y=-1+a-w, \quad {\rm and} \quad z=-1+b-w.\] Then using orthogonality between the second and third columns we get the equations (\ref{special}). Now since $|x|=|z|=1$ we have that $1+a \perp (1-a)b$ or $a^{\rm pu} \perp b^{\rm pu}$. Thus $b \in {\rm span}\{1,{\bf j}\}$ as required.

Suppose we now have a dephased quaternion Hadamard matrix having no real entries apart from those in the first row and column. By lemma \ref{column form II} the matrix is of the form:
\[ H = \begin{bmatrix}
1 & 1 & 1 & 1 \\
1 & a & b & c \\
1 & h-(1+a)/2 & k - (1+b)/2 & \star \\
1 & -h-(1+a)/2 & -k - (1+b)/2 & \star
\end{bmatrix} \]
where $|a|=|b|=|c|=1$; $2|h|=|1-a|$, $2|k|=|1-b|$ and $h$ and $k$ are orthogonal to $1+a$ and $1+b$ respectively. The orthogonality condition on the second and third columns gives \begin{equation}\label{ortho} -4\bar{h}k = (1+\bar{a})(1+b) + 2(1+\bar{a}b). \end{equation} As $h \perp 1+a$ and $k \perp 1+b$, we have \[ h=s(1+a), \quad k =t(1+b) \] where $s$ and $t$ are pure quaternions with \[ 2|s|=\left|\frac{1-a}{1+a}\right|, \quad 2|t|=\left|\frac{1-b}{1+b}\right|. \] Substituting into equation (\ref{ortho}) we get \begin{equation}\label{eq1} 4st = 1 + 2\frac{1+a}{|1+a|^2}(1+\bar{a}b)\frac{1+\bar b}{|1+b|^2}. \end{equation} Note that $c=-(1+a+b)$ so $|1+a+b|^2=1$ which is \[ 2+a+\bar a + b + \bar b + a \bar b + b \bar a = 0 \] and is equivalent to $1+a \perp 1+b$, and to \begin{equation}\label{perp} (1+a_1)(1+b_1)+a_2b_2+a_3b_3+a_4b_4=0. \end{equation} We have \begin{eqnarray*}(1+a)(1+\bar a b)(1+\bar b) & = & 2+a+\bar a + b + \bar b + a \bar b + \bar a b \\ & = & \bar a b - b \bar a.\end{eqnarray*} Now $|a|^2=|b|^2=1$ implies that $|1+a|^2|1+b|^2 = 4(1+a_1)(1+b_1)$. Since \[ a^{\rm pu}b^{\rm pu} = (a_1 - \bar a)(b-b_1) = -a_1b_1 + a_1b + b_1\bar a - \bar a b, \] \[ b^{\rm pu}a^{\rm pu} = (b - b_1)(a_1 - \bar a) = -a_1b_1 + a_1b + b_1 \bar a - b \bar a, \] and \[ a^{\rm pu}b^{\rm pu} + b^{\rm pu}a^{\rm pu} = -2 \langle a^{\rm pu}, b^{\rm pu} \rangle = 2(1+a_1)(1+b_1), \] we have \begin{eqnarray*} 2b^{\rm pu}a^{\rm pu} & = & \left( 2(1+a_1)(1+b_1)-a^{\rm pu}b^{\rm pu} \right) + b^{\rm pu}a^{\rm pu} \\ & = & 2(1+a_1)(1+b_1) + \bar a b - b \bar a. \end{eqnarray*} Hence (\ref{eq1}) can be written as \begin{equation}\label{stuv} st = uv \end{equation} where \[ u = \frac{b^{\rm pu}}{2(1+b_1)}, \quad v = \frac{a^{\rm pu}}{2(1+a_1)}. \] It is straightforward to check that $|s|=|v|$ and $|t|=|u|$. Equation (\ref{stuv}) is equivalent to \[ \langle s,t \rangle = \langle u,v \rangle, \quad s \times t = u \times v. \] Thus ${\rm span} \{ s,t \} = {\rm span} \{ u,v \}$ and the angle between $s$ and $t$ equals the angle between $u$ and $v$. Hence there is a rotation in the space of pure quaternions that takes $v$ to $s$ and $u$ to $t$, thus there is a unit norm quaternion $x$ such that \[ s=xv\bar{x}, \quad t=xu\bar{x}. \] Then (\ref{stuv}) gives that $x vu \bar x = uv$ which after rearranging becomes \begin{equation}\label{xyx} xyx^{-1} = \bar y \end{equation} where \[ y = \frac{vu}{|vu|} = \frac{a^{\rm pu}b^{\rm pu}}{|a^{\rm pu}b^{\rm pu}|}. \] If we plug into $H$ the expressions for $s$ and $t$ and switch the last two rows, we obtain the matrix:
\[ \begin{bmatrix}
1 & 1 & 1 & 1 \\
1 & a & b & \star \\
1 & -x \frac{1+a}{|1+a|} \bar x \frac{1+a}{|1+a|} & -x \frac{1+b}{|1+b|} \bar x \frac{1+b}{|1+b|} & \star \\
1 & \star & \star & \star
\end{bmatrix}. \] 

Now since we are free to conjugate the matrix entrywise by a unit norm quaternion, we may assume that $a = a_1 + a_2 {\bf i} + a_3 {\bf j}$ and $b = b_1 + b_2 {\bf i}$. Let $\hat{a} = a_1 + a_2 {\bf i}$. Since we assumed that $a, b \not \in \bR$, equation (\ref{perp}) implies that $a_2 \neq 0$ and so \[ b_2 = -\frac{(1+a_1)(1+b_1)}{a_2}. \] Then from $|b|^2=1$ we get \[ ((1+a_1)^2+a_2^2)b_1^2 + 2(1+a_1)^2b_1 + (1+a_1)^2 - a_2^2 = 0. \] One of the roots of this quadratic is $b_1=-1$, which is not the case. Thus we must have that \[ b_1 = - \frac{(1+a_1)^2 - a_2^2}{(1+a_1)^2 + a_2^2} \] and a simple computation gives \[ b = -\frac{(1 + \hat a)^2}{|1 + \hat a|^2}. \] If $a_3=0$, then $a$ is complex and the above formula  shows that $b=-a$ and $c=-1$ which gives a contradiction. We conclude that $a_3 \neq 0$. Since $a^{\rm pu}b^{\rm pu} = (a_2 {\bf i} + a_3 {\bf j}) b_2 {\bf i} = -b_2(a_2+a_3 {\bf k})$ and $a_3b_2 \neq 0$, the equation $xyx^{-1}=\bar y$ implies that $xkx^{-1}=-k$. Thus $x \in {\rm span}\{ {\bf i}, {\bf j}\}$, and in particular $\bar x = -x$. As \[ |1 + \hat a|^2 - (1 + \hat a)^2 = -2 a_2(1 + \hat a){\bf i}, \] we have \[ \frac{1+b}{|1+b|} = \pm \frac{1 + \hat a}{|1 + \hat a|} {\bf i}, \] and we arrive at the result.
\end{proof}

\section{Quaternionic MUB}\label{quat mub}

Two orthonormal bases $\{u_i\}_{i=1}^n$, $\{v_i\}_{i=1}^n$ in $\bC^n$ are said to be {\em mutually unbiased bases} (MUB) if \[ | \langle u_i,v_j \rangle |^2=\frac{1}{n} \] for all $1 \leq i,j \leq n$. A collection of orthonormal bases for $\bC^n$ is said to be a set of mutually unbiased bases if they are pairwise mutually unbiased. We extend this definition in the natural way to $\bH^n$. We consider $\bH^n$ as the right vector space over $\bH$ consisting of column vectors. We use the physicists' notation $ | x \rangle $ to denote a column vector and $ \langle x |$ to denote the conjugate transpose vector $| x \rangle ^ *$. The space $\bH^n$ is equipped with the standard positive definite inner product $\langle x | y \rangle = | x \rangle^* | y \rangle$. We identify an orthonormal basis with any symplectic matrix that has as columns the vectors in the basis. The condition that a set $S=\{A_1, \ldots ,A_k\}$ of symplectic matrices be mutually unbiased is simply that $\sqrt n A_i^*A_j$ is Hadamard for all $1 \leq i < j \leq k$. If a set of MUB contains only Hadamard matrices (more precisely, matrices $A_i$ such that $\sqrt n A_i$ is Hadamard), we may include the identity matrix to create another set of MUB. Conversely if $\{A_1, \ldots ,A_k\}$ is a set of MUB, then $\{I,A_1^*A_2, \ldots ,A_1^*A_k\}$ is a set of MUB, where $I$ is the $n$ by $n$ identity matrix and $\sqrt n A_1^*A_j$ is Hadamard. Two sets of MUB $\{A_0,A_1, \ldots ,A_k\},\{B_0,B_1, \ldots ,B_k\}$ are to be considered {\em equivalent} if there exist symplectic monomial matrices $E_i$ for $0 \leq i \leq k$ and a symplectic matrix $U$ such that $UA_iE_i=B_{\pi(i)}$ for all $0 \leq i \leq k$ where $\pi$ is a permutation of $\{0, \ldots ,k\}$. If a set of MUB contains the identity matrix, we may then additionally dephase some other matrix in the set.

It is known that the upper bound on the size of a set of MUB in $\bC^n$ is $n+1$ \cite{BBRV}. For $\bH^n$ this is not the case, see Theorem \ref{quat mub n=2}. 

We give an upper bound:

\begin{theorem} For $n \geq 2$, a set of quaternionic MUB in $\bH^n$ consists of at most $2n+1$ bases. \end{theorem}
\begin{proof}
Let $\bH_n$ be the real vector space of all Hermitian quaternionic $n$ by $n$ matrices, and $\bH_n^0$ its subspace consisting of zero trace matrices. Note that $\dim \bH_n^0 = 2n^2-n-1 = (n-1)(2n+1)$. The inner product in $\bH_n$ is given by \[ \langle A,B \rangle = \Tr (AB), \] where for any $n$ by $n$ quaternionic matrix $X$ we define its trace by \[ \Tr X = 2 \sum_{i=1}^n {\rm Re}(X_{ii}). \] Let $\pE=\{| e_1 \rangle, ..., | e_n \rangle\}$ be an orthonormal basis of $\bH^n$. To each unit vector $| e \rangle \in \bH^n$ we associate the operator \[ | e \rangle \langle e | - \frac{1}{n}I_n \in \bH_n^0. \] To the basis $\pE$ we associate the $(n-1)$-dimensional subspace $V_\pE \subseteq \bH_n^0$ spanned by the matrices \[ E_i = | e_i \rangle \langle e_i | - \frac{1}{n}I_n, \ 1 \leq i \leq n. \] Observe that the sum of these matrices is $0$ and they are vertices of a regular $(n-1)$-simplex in $\bH_n^0$. Let $\pF=\{f_1,...,f_n\}$ be another orthonormal basis of $\bH^n$ and assume that $\pE$ and $\pF$ are mutually unbiased. If \[ F_j = | f_j \rangle \langle f_j | - \frac{1}{n}I_n, \ 1 \leq j \leq n, \] then we have \begin{eqnarray*} \langle E_i, F_j \rangle & = & \Tr (E_iF_j) \\ & = & \Tr(| e_i \rangle \langle e_i | f_j \rangle \langle f_j | ) - \frac{1}{n}\Tr( | e_i \rangle \langle e_i | ) - \frac{1}{n}\Tr( | f_j \rangle \langle f_j | ) + \frac{2}{n} \\ & = & 2 \left| \langle e_i | f_j \rangle \right|^2 - \frac{2}{n} \\ & = & 0. \end{eqnarray*} Consequently $V_\pE \perp V_\pF$. Since each of $V_\pE$, $V_\pF$, $\ldots$ has dimension $n-1$ and $ \dim \bH_n^0 = (n-1)(2n+1)$, the assertion of the theorem follows.
\end{proof}

\begin{remark}
This proof is an obvious adaptation of the proof in the complex case sketched in the paper \cite{BBELTZ}.
\end{remark}

A set of $2n+1$ MUB in $\bH^n$ will be called {\em complete}. A set of MUB is {\em maximal} if it cannot be extended to a larger MUB. We shall see later that there exist maximal MUB which are not complete. Next we show that the above bound is attained in the case $n=2$.

\begin{theorem}\label{quat mub n=2}
For $n=2$ there exists a complete set of MUB in $\bH^n$. Moreover, it is unique up to equivalence. The matrices
\[ \begin{bmatrix}
1 & 1 \\
1 & -1 \\
\end{bmatrix}, \quad 
\begin{bmatrix}
1 & 1 \\
{\bf i} & -{\bf i} \\
\end{bmatrix}, \quad
\begin{bmatrix}
1 & 1 \\
{\bf j} & -{\bf j} \\
\end{bmatrix}, \quad
\begin{bmatrix}
1 & 1 \\
{\bf k}& -{\bf k}\\
\end{bmatrix},\]
multiplied by $1/ \sqrt 2$ and the identity matrix form such an MUB.
\end{theorem}
\begin{proof}
Assume $\{I,A_1,...,A_4\}$ is a set of MUB. Thus $\sqrt 2 A_i$ and $\sqrt 2 A_i^*A_j$, $i \neq j$, are $2$ by $2$ Hadamard matrices. After dephasing we may assume
\[ \sqrt 2 A_1 = \begin{bmatrix}
1 & 1 \\
1 & -1 \\
\end{bmatrix}. \]
If $n \geq 2$ we multiply $A_2$ on the right by an appropriate symplectic diagonal matrix to get
\[ \sqrt 2 A_2 = \begin{bmatrix}
1 & 1 \\
a & -a \\
\end{bmatrix}, \]
for some unit norm quaternion $a$. Then since $\sqrt 2 A_1^*A_2$ is Hadamard we have that $|1+a|=|1-a|=1$ so $a$ is pure. We may conjugate by $x \in \bH$ to make $a={\bf i}$. Next we have
\[ \sqrt 2 A_3 = \begin{bmatrix}
1 & 1 \\
b & -b \\
\end{bmatrix}, \]
for some unit norm quaternion $b$. Again $b$ must be pure but also we see that ${\bf i}b$ is pure. Thus $b \in {\rm span} \{ {\bf j},{\bf k}\}$. Conjugating by a complex number we have $b={\bf j}$. Finally, $A_4$ has a similar form involving a unit norm quaternion $c$. Now we have $c$, ${\bf i}c$ and ${\bf j}c$ must be pure. This means $c={\bf \pm k}$ and if necessary we may switch the two columns of $A_4$ to have $c={\bf k}$. A computation shows this is indeed a set of MUB.
\end{proof}

Now we consider the problem with $n=3$. Let us introduce six families of quaternionic $3$ by $3$ Hadamard matrices which are unbiased with respect to the Fourier matrix $F_3$. All these claims are easy to verify by using Maple. All of the families consist of matrices with the same basic form 
\begin{equation}\label{mub3} \begin{bmatrix}
1 & 1 & 1 \\
a & a\zeta & a\zeta^2 \\
b & b\zeta^2 & b\zeta \\
\end{bmatrix}, \end{equation} where $|a|=|b|=1$ and \[ \zeta = -\frac{1}{2} + s {\bf i} + t {\bf j}, \quad s^2+t^2=3/4; \quad s,t \in \bR. \] Let \begin{equation} p(a,s,t) = (a_3^2 + a_4^2)s + (a_1a_4 - a_2a_3)t \end{equation} and \begin{equation}\label{phi} \varphi(a,s,t) = 4 \alpha_0 s^2 + 8 \alpha_1 st + \alpha_2 \end{equation} where \begin{equation}\label{alphas} \begin{array}{lcl} \alpha_0 & = & 1-a_1+4a_1a_2^2+2a_1a_4^2+2a_2a_3a_4-2a_3^2-2a_4^2, \\ \alpha_1 & = & a_1^2a_4-a_2^2a_4+2a_1a_2a_3-a_1a_4+a_2a_3 , \\ \alpha_2 & = & 1-a_1+4a_1a_2^2+4a_1a_3^2-2a_1a_4^2-6a_2a_3a_4. \end{array} \end{equation}

First there is a $3$-parameter family, to which we refer as the {\em generic family}, consisting of matrices (\ref{mub3}) where $\varphi(a,s,t)=0$, $p(a,s,t) \neq 0$, and $b$ is the unique solution to \begin{equation}\label{sys} \langle 1+\omega^{-i} a \zeta^j , 1 + \omega^i b \zeta^{-j} \rangle = 1, \quad i,j \in \{0,1\}. \end{equation}

There are also five {\em special families}, each depending on two parameters. It should be understood that, in each case, only the listed restrictions apply in addition to $|a|=|b|=|\zeta|=1$.

\begin{enumerate}[1:]
\item $a=1$, $b_1 = -\frac{1}{2}$ and $b_4=0$.

\item $a = \zeta$, $\zeta^2$; $b_1 = 1 - 2(a_2b_2+a_3b_3)$ and $b_4 = 2(a_3b_2 - a_2b_3)$.

\item $a = \omega$, $\omega^2$; $b_1 = 1 - 2a_2b_2$ and $b_4 = 2a_2b_3$.

\item $a_1=\frac{1}{4}$, $a_2^2=a_3^2=\frac{3}{16}$, $a_4=4a_2 a_3$, $b_1 = -\frac{1}{2} - \frac{4}{3}a_4b_4$, $b_3 = \frac{2a_3}{3}(1-4b_1-8a_2b_2)$ and $s=0$.

\item $3a_2^2 = (1-a_1)^2$, $6a_3^2=(1-a_1)(1+2a_1)$, $a_4^2=3a_3^2$, $p(a,s,t)=0$, $b_1 = -\frac{1}{2} - \frac{a_2b_4}{a_3}$, $b_2 = \frac{1-a_1}{2a_2}-\frac{a_3b_3}{a_2}+\frac{b_4}{2a_3}$ and $t^2 = 4a_3^2$.

\end{enumerate}

The next theorem reveals the importance of these families for the problem of constructing a maximal set of MUB. We shall give two examples at the end of this section.

\begin{theorem}\label{qmub3}
Let $n=3$. Any quaternionic MUB $\{ A_0, A_1, A_2 \}$ is equivalent to one where $A_0 = I_3$, $A_1 = F_3$ is the Fourier matrix, and $A = \sqrt 3 A_2$ belongs to one of the six families defined above.
\end{theorem}
\begin{proof}

We may assume that $A_0 = I_3$. Since any $3$ by $3$ quaternion Hadamard is equivalent to $\sqrt 3 F_3$, we may assume $A_1 = F_3$. We may also assume that the Hadamard matrix $A$ has the form
\[ A = \begin{bmatrix}
1 & 1 & 1 \\
a & c & e \\
b & d & f \\
\end{bmatrix}. \]
Let $x$ be a unit quaternion such that $x^{-1}a^{-1}cx$ is complex, let $X = {\rm diag} (x,x,x)$ and let 
\[ D =
\begin{bmatrix} 1 & 0 & 0 \\
0 & a^{-1} & 0 \\
0 & 0 & b^{-1} 
\end{bmatrix}.\] Then we have that
\[ X^{-1} D A X =
\begin{bmatrix}
1 & 1 & 1 \\
1 & x^{-1}a^{-1}cx & x^{-1}a^{-1}ex \\
1 & x^{-1}b^{-1}dx & x^{-1}b^{-1}fx \\
\end{bmatrix} \]
is Hadamard and by orthogonality we see that all entries are complex. By having permuted the columns of $A$ if necessary, we see that the RHS of this equation must be $\sqrt 3 F_3$. Then we have $A = \sqrt 3 D^{-1} X F_3 X^{-1}$. Thus if we let $\zeta = x \omega x^{-1}$ we get that $A$ is in the form (\ref{mub3}).

Now since $A_2$ must be unbiased with respect to $F_3$ we have that equations (\ref{sys}) hold. If we take $a$ and $\zeta$ to be fixed, we obtain a linear system of equations $Bb=v$ where $B$ is a $4$ by $4$ real matrix and $v$ is a real column vector. The entries of $B$ and $v$ depend on $a$ and $\zeta$. Note that since $I_3$ and $F_3$ are complex, we are free to conjugate $A$ entrywise by a complex number of unit norm, and thus we assume that $\zeta \perp {\bf k}$.

Let $d_i$ be the determinant of the $4$ by $4$ matrix obtained by dropping the $i^{\rm th}$ column of the augmented matrix $\tilde{B}=[B|v]$.  It can be checked, e.g. with Maple, that \begin{equation}\label{id1} d_5 \equiv 3 p(a,s,t)^2, \end{equation} and \begin{equation}\label{id2} 8 \left( \sum_{i=1}^4 d_i^2 - d_5^2 \right) \equiv 9 d_5 \varphi (a,s,t) \end{equation} where the congruence is modulo the ideal $\langle |a|^2-1, |\zeta|^2-1 \rangle$.

Now if ${\rm det} B = d_5 \neq 0$ then by Cramer's rule we have $b_i = \pm d_i / d_5$ and since $|b|=1$ we get that $\sum_{i=1}^4 d_i^2 = d_5^2$. From (\ref{id1}) and (\ref{id2}) it follows that $p \neq 0$ and $\varphi = 0$. Thus $A$ is in the generic family.

If $d_5=0$ then (\ref{id1}) implies $p(a,s,t)=0$. Note that by (\ref{id2}) we have then that $d_i=0$ for $1 \leq i \leq 4$. We now distinguish three cases.

Case $1$: $a_3 = a_4 = 0$. Rows $1$ and $3$ of $\tilde{B}$ reduce to the system \[ (1 + a_1)b_1 + a_2b_2 = -a_1, \] \[ a_2b_1 + (1-a_1)b_2 = a_2.\] Since \[ \left| \begin{matrix} 1+a_1 & a_2 \\ a_2 & 1-a_1 \end{matrix} \right| = 0, \] the two equations must be linearly dependent and so $a_1-a_1^2+a_2^2 = a_2(1+2a_1) = 0$. Thus either $a_2 = 0$ and so $a = 1$, or $a_1 = -\frac{1}{2}$ and $a = \omega$  or $\omega^2$. If $a=1$ then we see that $b_1=-\frac{1}{2}$ and row $4$ gives either $t=0$ or $b_4=0$. If $t=0$ then $\zeta$ is complex and by conjugating $A$ entrywise by a complex number we can assume $b_4=0$. Thus we get that $A$ is in Family $1$. If $a_1=-\frac{1}{2}$ then we see that by rows $1$ and $2$ we have $b_1 + 2 a_2 b_2 = 1$ and $t(3b_3 - 2a_2b_4) = 0$. Thus we get that $t = 0$ or $b_4 = 2 a_2 b_3$. If $b_4 = 2 a_2 b_3$ then $A$ is in Family $3$. If $t=0$ then $\zeta$ is complex and by conjugating $A$ entrywise with a complex number we see that $A$ is in Family $2$.

Case $2$: $a_3^2 + a_4^2 > 0$; $a_1a_4 - a_2a_3 = 0$. Then $s=0$. By calculating a Groebner basis for the ideal $\langle s, a_1a_4-a_2a_3,|\zeta|^2-1, |a|^2-1 \rangle$ one may check that all $3$ by $3$ minors of $B$ are $0$. Thus all $3$ by $3$ minors of $\tilde{B}$ have to be $0$ as well. After computing a Groebner basis for the ideal generated by the generators of the previous ideal, along with the remaining minor equations, one sees that $a_4 \in \{ 0, \pm \frac{3}{4} \}$. If $a_4=0$ then $a_3 \neq 0 $ so $a_2=0$. Clearly $a_1 \neq 1$ since $a_3^2 + a_4^2 > 0$, thus we see that $a_1=-\frac{1}{2}$. Solving the system we get $b_1 = 1 - 2a_3b_3$ and $b_4 = 2a_3b_2$ thus $A$ is in Family $2$. If $a_4=\pm \frac{3}{4}$ then $a_1=\frac{1}{4}$ and $a_3^2=\frac{3}{16}$. Furthermore $a_2=\frac{4}{3}a_3a_4$. Then solving the system we find from row $1$ that $b_1 = -\frac{1}{2} - \frac{4}{3}a_4b_4$ and then $b_3 = \frac{2a_3}{3}(1-4b_1-8a_2b_2)$. Thus $A$ is in Family $4$.

Case $3$: $a_3^2 + a_4^2 > 0$; $a_1a_4 - a_2a_3 \neq 0$. Then $s \neq 0 $ and $t \neq 0$. By calculating a Groebner basis for the ideal $\langle p(a,s,t), |\zeta|^2-1, |a|^2-1 \rangle$ one may check that all $3$ by $3$ minors of $B$ are $0$. Thus all $3$ by $3$ minors of $\tilde{B}$ have to be $0$ as well. Collect all $3$ by $3$ minors of $\tilde{B}$ and append the polynomials $|\zeta|^2-1$, $|a|^2-1$ and $p(a,s,t)$, and compute the Groebner basis of the ideal $J$ they generate. By using this basis it is easy to verify that the set of polynomials \[ \{ a_4 t (3a_3^2-a_4^2), a_3 (t^2-a_3^2-a_4^2), t (3a_3^2 - a_4^2)(1+2a_1), \] \[4a_3^2 (st-a_2 a_3)+a_4 (a_3^2+a_4^2-4a_2 a_3 a_4), a_4 t (a_2 a_4 - a_3 (1-a_1)), \] \[t(a_2(1+2a_1) - 2a_3 a_4) \}, \] is contained in $J$. From $a_4 t (3a_3^2-a_4^2)=0$ we deduce that $a_3 \neq 0$ and that either $a_4 = 0$ or $a_4^2 = 3 a_3^2$. If $a_4=0$ we obtain that $a_1 = -\frac{1}{2}$, $st=a_2 a_3$, and $t^2 = a_3^2$. Thus $s^2 = a_2^2$ and $a = \zeta$ or $\zeta^2$. In both cases we get $b_1 = 1 - 2(a_2b_2+a_3b_3)$ and $b_4 = 2(a_3b_2 - a_2b_3)$. Thus $A$ is in Family $2$. If $a_4 \neq 0 $ then $a_4^2 = 3a_3^2$, $t^2 = 4a_3^2$, $a_2 a_4 = a_3 (1-a_1)$ and $2a_3 a_4 = a_2(1+2a_1)$. From the last two equations we find that $2a_4^2 = (1-a_1)(1+2a_1)$ and \[ 3a_2^2(1+2a_1)=6a_3^2(1-a_1)=2a_4^2(1-a_1)=(1-a_1)^2(1+2a_1), \] which gives $3a_2^2=(1-a_1)^2$. Now rows $1$ and $2$ of system $\tilde{B}$ give \[ b_1 = -\frac{1}{2} - \frac{a_2b_4}{a_3}, \quad b_2 = \frac{1-a_1}{2a_2}-\frac{a_3b_3}{a_2}+\frac{b_4}{2a_3}.\] Thus $A$ is in Family $5$.
\end{proof}

As promised, we now give two families of genuine quaternionic MUB consisting of four bases (i.e., symplectic matrices). The first one depends on one parameter only, while the second depends on three parameters.

\begin{example}
We give here a one parameter family of quaternionic MUB of size $4$ which includes the complete complex MUB as a particular case. They are parameterized by the points $(s,t)$ on the circle $s^2+t^2=3/4$ and are given by \[ I_3, F_3, \frac{1}{\sqrt 3} A(s,t), \frac{1}{\sqrt 3} A(-s,-t) \] where 
\[ A(s,t) = \begin{bmatrix}
\zeta & 1 & 1 \\
1 & \zeta & 1 \\
1 & 1 & \zeta
\end{bmatrix}, \quad \zeta = - \frac{1}{2} + s {\bf i} + t {\bf j}. \] For $t=0$, $s = \sqrt 3 / 2$ this is a complex MUB. 

We have verified using Maple and Theorem \ref{qmub3}, that such MUB, for arbitrary $(s,t)$ is not extendible to an MUB of size $5$.
\end{example}

\begin{example}
We now give a normalized $3$-parameter family of maximal quaternionic MUB: \[ I_3, F_3, \frac{1}{\sqrt 3}A(a,b,c), \frac{1}{\sqrt 3}B(a,b,c). \] The $a$, $b$, $c$ are any quaternionic cubic roots of unity orthogonal to ${\bf k}$. Thus each of them has the form \[ -\frac{1}{2}+s{\bf i} + t{\bf j}; \quad s^2 + t^2 = 3/4. \] The Hadamard matrices $A$ and $B$ belong to our family $1$ and are given by 
\[ A(a,b,c) = 
\begin{bmatrix}
1 & 1 & 1 \\
1 & a & a^2 \\
b & ba^2 & ba
\end{bmatrix}, \quad 
B(a,b,c) = 
\begin{bmatrix}
1 & 1 & 1 \\
1 & c & c^2 \\
\bar b & \bar bc^2 & \bar bc
\end{bmatrix}.
\] The family is normalized in the sense that we have fixed the first two matrices to be $I_3$ and $F_3$.

It is not hard to verify that the four orthonormal bases provided by the columns of these matrices are indeed MUB. For example if we take the conjugate transpose of the second column in $A$ and multiply by the second column of $B$ we get the sum $1 + a^2c + abc^2$. By multiplying on the left and right by $a^2$ and $c$ respectively, this sum has the same norm as $a^2c+ac^2+b = 2(a_2c_2 + a_3c_3) + b_2 {\bf i} + b_3 {\bf j} + 2(a_3c_2 - a_2c_3){\bf k}$. The norm squared of this sum is easily seen to be $3$ as required. If we take the special case $a=b=c=\zeta$, we obtain a $1$-parameter family which is equivalent to the one from the previous example.
\end{example}

\section{Open Problems}\label{openp}

Here we mention some relevant problems that remain open.

\begin{problem} For arbitrary dimension $n$, real, complex or quaternionic, characterize the singular points of $\phi_r : \Ort(n) \rightarrow \pA_n$ and the critical points of $\phi_c : \U(n) \rightarrow \pA_n$ and $\phi_h : \Sp(n) \rightarrow \pA_n$.

The singular points of $\phi_r$ for $n \leq 4$ only depend on the positions of zero entries in the orthogonal matrix. It would be interesting to see if this phenomenon persists in higher dimensions.
\end{problem}

\begin{problem}
Is it true that the twelve polynomial equations $\Sigma_4$ satisfied by $\pO_4$ characterize this set?
\end{problem}

\begin{problem}
Determine all maximal sets of MUB in $\bH^3$.

We have found a $3$-parameter family of such MUB consisting of four bases. This family contains the maximal set of complex MUB that is unique up to equivalence in $\bC^3$.
\end{problem}

\begin{problem}
Describe the set of all quaternionic Hadamard matrices of size $5$ or $6$. 

We have found two families of quaternionic Hadamard matrices of size $4$ and we have shown that any quaternionic Hadamard matrix of size $4$ is equivalent to a matrix in one of these families.
\end{problem}

\begin{problem}\label{sing}
Prove that the maps $\phi_c : \U(n) \rightarrow \pA_n$ and $\phi_h : \Sp(n) \rightarrow \pA_n$ have regular points for all $n \geq 1$. 

The assertion is easy to verify for small values of $n$ and it is known to be true if $n$ is prime \cite{TZS}. This problem is mentioned in \cite{BEKTZ}.
\end{problem}

\end{document}